\documentclass[epj-spec]{svjour}
\usepackage{graphics}
\usepackage{epsfig}

\newcommand{\be}{\begin{equation}}
\newcommand{\ee}{\end{equation}}
\newcommand{\bea}{\begin{eqnarray}}
\newcommand{\eea}{\end{eqnarray}}
\begin{document}

\title{Separating a mixture of chaotic signals\footnote{This paper was presented at International Conference on Non-linear Dynamics and Chaos:
Advances and Perspectives, 17--21 September 2007, Aberdeen, UK.}}


\author{Prabhakar G. Vaidya\thanks{\email{pgvaidya@yahoo.com}}}

\institute{National Institute of Advanced Studies, Indian Institute
of Science Campus, Bangalore 560 012, INDIA}

\abstract{ Chaos is popularly associated with its property of  sensitivity to initial conditions. In this paper  we will show that there can be a flip side to this property which is quite fascinating and highly useful in many applications. As a result, we can mix a large number of chaotic signals and one completely arbitrary signal and  later a recipient of this transformed and weighted mixture can separate each of the signals, one by one. The chaotic signals, could be generated by various maps which belong to the logistic family. The arbitrary signal, could be a message, some random noise, some periodic signal or a chaotic signal generated by a source, either belonging or not belonging to the family.   The key behind this procedure is a family of maps which can dovetail into each other without altering each of their predecessor's symbolic sequence.}

\maketitle

\section{Introduction}

Chaos is popularly associated with its property of sensitivity to
initial conditions. In this paper  we will show that there can be  a
flip side to this property.  This side could be found in the case of
maps which generate chaotic trajectories. These maps are surjective
but not bijective. Therefore there is usually no unambiguous way to
go backwards in time. However, if we happen to know the symbolic
sequence, reversal is unambiguous. What is more, the conditional
reversal maps are contractions and therefore forgiving of errors.
This property can be harnessed so that we can mix a large number of
chaotic signals and one completely arbitrary signal and  later a
recipient of this transformed and weighted mixture can separate each
of the signals, one by one. In this paper, we will assume that the
chaotic signals are generated by various maps which belong to the
logistic family. The arbitrary signal, could be a message, some
random noise, some periodic signal or a chaotic signal generated by
a source, either belonging or not belonging to the family.

 The idea behind this study arose soon after a casual discussion in which  Professor Walter Freeman~\cite{Walter} made a
strong case that  the EEG signals are chaotic. Now, if the brain is
the command and communication center of the body how could it
operate with ``chaotic" signals? How do you keep track of which
signal came from where?

This paper will now show that the chaotic nature of the signals,
instead of being a handicap, is indeed an asset. In fact, today the
main motivation for  the study of this problem seems be the fact
that chaotic maps could be applied  to the areas of cryptography,
communication and error correction codes (well documented by my
colleague Nithin Nagaraj~\cite{Nithin1,Nithin2,Nithin3}). This study
has a potential for applications in all these three areas.

In what follows, there is, at first, a review of how  various maps
can be related using topological conjugacy. Then there is a
description of  how to create a single  noise resistant map. This is
followed by `a procedure  to  create a separable cascade of  maps
which dovetail into each other without altering each of their
predecessor's symbolic sequence. The orbits of these maps can then
be mixed at the sender's end and disentangled at the receiver's end
by  using symbolic sequences.

This is followed by a theoretical and numerical demonstration, which
begins with $M$ chaotic signals, all generated by the standard tent
map, with a random set of initial conditions each, producing $M$
independent chaotic signals of length $L$ each. We then include one
more member in this collection which is also of a length $L$. This
last signal is quite arbitrary. It could be a message, some random
noise, some periodic signal or a chaotic signal generated by a
source, either belonging or not belonging to the family. The first
$M$ signals naturally have all their terms ranging from 0 to 1. For
the last signal, an affine transformation (which is obviously
invertible) might be needed to bring its terms within the same
range. All the $M+1$ signals are now treated as if they all belonged
to the logistic (or the standard tent map) family and each of them
is transformed into a different noise resistant map to once again
form a separable cascade. The trajectories from these are added and
then using a program simulating the receiver, separated and
reconstructed into the standard tent map setting. Based on the
accuracy specified, a portion of length $Q$ of the tail each signal
is abandoned but the $M+1$ sequences of the reduced length $L-Q$
each are highly accurately reproduced by a single message of length
$L$.

\begin{figure}[!h]
\centering
\includegraphics[scale=0.6]{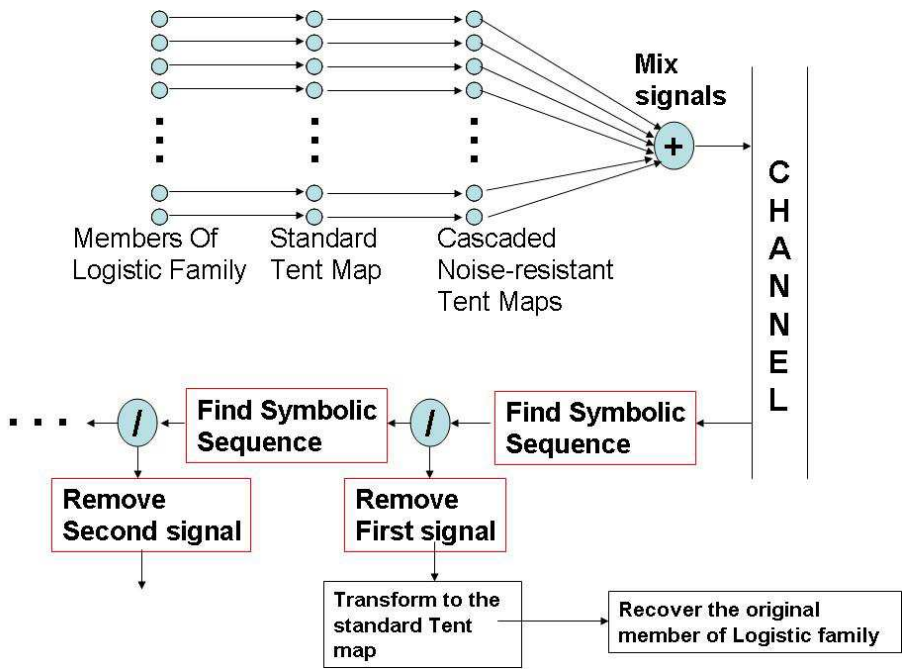}%
\caption[Figure 1]{Separating a mixture of chaotic signals.}
\label{figure:fig1}
\end{figure}

 In a section after the numerical simulation we discuss a scenario which consists of many signals from possibly
different maps, (although all belonging to the same conjugate
family), coming to a central mixing station.  (Please see
Figure~\ref{figure:fig1}). At this station, each of these signals
are transformed to the standard tent map and sent to a receiving
station and the receiver upon receiving follows a reversal procedure
by which she recovers all the signals with a remarkable accuracy.

\section{A review of Topological Conjugacy and its use to generate maps which share the same symbolic sequence}

Consider a map $U$ that takes $x$ to $U(x)$, that is $x_{n+1}=
U(x_{n})$. In order to create a map which is topologically conjugate
to this map, we can begin by choosing a   function $f$ ,which is
continuous with continuous inverse,  and using this   define $y =
f(x)$. From this we can  create a new map $V$ which  takes $y$ to
$V(y)$ by defining $V(y)= f(U(f^{-1}(y)))$. $U$ and $V$ are said to
be topologically conjugate.

Such maps share a lot of interesting properties. Of importance for
this paper are these two. The first is self evident. If the initial
condition of the first map is $x_{0}$, we know that for the second
map, the corresponding initial condition would be $y_{0}= f(x_{0})$
and if the first map generates an orbit $x_{0},x_{1},x_{2}...$ we
can generate the  corresponding orbit $y_{0},y_{1},y_{2},..$  by two
different ways: One is to iteratively use $V$ and the other is to
operate on the $x$ sequence by $f$. The second property is the
preservation of symbolic sequences. Let the domain for $U$ be
divided into suitable partitions, known as Markov Partitions, (for
the standard  tent map below, we choose the symbol 0 if $x$ falls in
[0,0.5) and 1 if it falls in [0.5,1] ). Operating $f$ on each of the
partitions bijectively maps it  into a corresponding partition of
$y$. As a result  the symbolic sequences for $x$ and $y$ will be
identical.

There is a large family of conjugate maps which has the logistic map
as one of its most well known  members. This family has another key
member  known as the standard tent map (see below).  We would assume
in the rest of the paper that the signals to be mixed by the sender
are members of this family. We achieve the goal of this paper by
creating a new set of members of this family which are parametrized
by two parameters $p$ and $q$ and all the members of this new
subfamily share an interesting property, described in the next
section.

\section{Noise resistant sub-family of symbolically equivalent maps}

We begin with the standard tent map from [0,1] to [0,1], $T(x):$

$$
= 2x,  ~~~  x \in [0,0.5),
$$
$$
    = 2-2x, ~~~ x \in [0.5,1].
$$

\begin{figure}[!h]
\centering
\includegraphics[scale=.6]{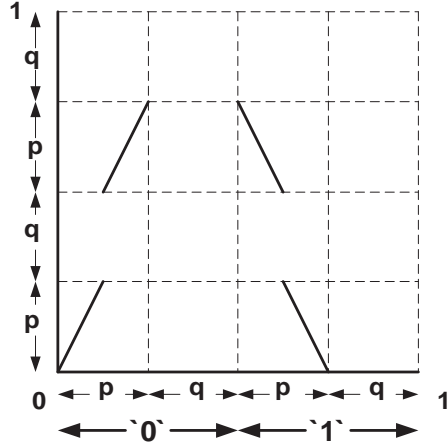}
\caption[Method 1: Noise-resistant Tent map]{Noise-resistant Tent
Map.} \label{fig:fignoiseresmap}
\end{figure}

To generate a sub family of maps (see
Figure~\ref{fig:fignoiseresmap}) symbolically equivalent to it we
choose the transformation $f$ as $y=f(x,p,q)$:

              $$
                  = 2px, ~~~ x\in[0,0.5),
              $$
              $$
                = 2px+q, ~~~ x\in[0.5,1].
               $$

\noindent which leads to a domain for $y$ given by the union of
$[0,p)$ and $[p+q, 2p+q]$ and we have an inverse transformation $
x=f^{-1}(y,p,q)$:
                                  $$
                                                    =y/2p, ~~~ y
                                                    \in[0,p),
                                                  $$
                                                  $$
                                                    =(y-q)/2p, ~~~ y
                                                    \in[p+q,2p+q].
                                                   $$

Using these two, we get a map $N(y)$ for the variable $y$ from the
union of $[0,p)$ and $[p+q, 2p+q]$ to itself $N(y)= f(T(f^{-1}(y)))$

This can be explicitly expanded as $N(y):$
$$
= 2y ~~ y \in [0,p/2),
$$

$$
= 2y+q, ~~ y \in [p/2,p],
$$

$$
=4p+3q-2y, ~~ y \in [p+q,(3p+2)/q],
$$

$$
=4p+2q-2y, ~~ y \in ((3p+2)/q,2p+q].
$$

Thus the two parameters $p$ and $q$ create a whole subfamily of maps
which are topologically conjugate and therefore symbolically
equivalent to the standard tent map. For a specific $p$ and $q$ we
can define a symbolic sequence corresponding a trajectory generated
by $N(y)$ by using the symbol 0 if $x$ is less than $p+q$ and 1
otherwise. This definition helps us embed the domain of the map in a
larger interval $[0,2p+2q]$. This extended domain becomes useful
when we define an additional variable $z = y +$ an additional
signal. Now if the additional signal is non-negative and of a
magnitude less than $q$ then it is clear that no matter what $y$ is
$z$ will remain in the extended domain.

We will soon make use of the fact that,  given a sequence of $y$ and
corresponding $z$ , the symbolic sequence of $y$ and $z$ would
remain identical.

\section{Recovery of signal from a symbolic sequence, in absence of noise }

Supposing we start with an initial condition and generate an orbit
and also a corresponding symbolic sequence. To what extent can we
reconstruct the orbit from the symbolic sequence alone? We would
answer this question, first in the absence of noise and then in
presence noise or some external signal.

First note that for any allowed combination of p and q the map is
surjective but not injective. However, if we are privy to the
symbolic sequence, the map becomes invertible. In fact, we can see
that $N^{-1}(y,p,q,S)$
$$
= y/2 ~~~if~~~ (S=0)\wedge (y<p+(q/2)),
 $$
$$
= (y-q)/2  ~~~if~~~ (S=0)\wedge (y\geq p+(q/2)),
$$
$$
=2p+q-(y/2)  ~~~if~~~  (S=1)\wedge (y<p+(q/2)),
$$
$$
=2p+(3q)/2-(y/2)  ~~~if~~~  (S=1)\wedge(y\geq(p+(q/2)).
$$

where $S$ is the symbolic sequence entry of the previous point in
the orbit.

Please note also that this inverse map contracts any interval to one
or two intervals with (total) length of half the  original length.
So, even if we do not know the end point of the orbit, we could
assume two extreme initial conditions and back iterate and after $Q$
steps the orbit is ``localized" to a measure of $2^{-Q}$ times the
original measure.

Thus the earlier a point in the orbit is, the more accurately is its
position determined by the inverse map. This also tells us that if
we need to determine all the points of an orbit within an acceptable
error of $\epsilon$  for a length of orbit $N$,  then we need to run
the original calculation for  $N+Q$ iterations and supply the
symbolic sequence of the whole orbit, where $Q$ has be greater than
$-\log_2(\epsilon)$.

It should be noted that there is a computationally more efficient
way to find the orbit  because the family of maps is topologically
conjugate to the standard tent map. This implies that for every
point on the map there is a corresponding point on the standard map
(given by $f^{-1}$ above) which follows an identical symbolic
sequence. For the standard map finding an initial condition can be
done by reading  the symbolic sequence from left to right and
repeating it if an even number of 1's have preceded in the original
sequence so far, and ``flipping" it    (from 0 to 1 or from 1 to 0)
if an odd number of 1's have preceded. The new sequence represents
location of the initial condition in binary~\cite{PGVaidya}.

\section{Noise resistance}

If we add to a sequence $y_{1},y_{2},y_{3}...$ some external signal
or noise sequence $r_{1},r_{2},r_{3}...$ and get a new sequence
$z_{1},z_{2},z_{3}...$ it is clear that $y$ and $z$ share the same
symbolic  sequence if $r_{1}$ etc. are non-negative and have a
magnitude less than $q$.

Therefore from the symbolic sequence of $z$ we can determine the
orbit of $y$ in the same manner and within the same accuracy, as if
the noise had no effect on it at all. Thus the entire family of maps
is resistant to noise up to  a magnitude equal to the parameter $q$.

\section{Cascading}

We begin with specific values for $p$ and $q$, say $p_{1}$ and
$q_{1}$. We select a second set $p_{2}$ and $q_{2}$ so that
$q_{1}\geq 2(p_{2}+q_{2}$). Now if we take any two orbits generated
by these two maps and add them, it is clear that the sum follows the
symbolic sequence of only the first one. Yet, the sum has retained
information about both of the sequences because using the symbolic
sequence we can determine the first orbit to a predetermined
accuracy  and then subtract it to get the second orbit.

This process can be continued for a fairly long cascade, each using
the noise tolerant corridor of length $q_{n}$  to accommodate the
extended diameter of the next stage 2($p_{n+1}+q_{n+1}$) . From a
sum of all these signals, starting from the first one all the
signals can be recovered one by one, provided we provide sufficient
additional length (e. g. $Q$ above) so that the errors in recovery
can be regarded as acceptable at each stage.

\section{Sender and Receiver}

In the next section we will describe numerical results of a
simulation in which there is a ``sender'' who first generates  $M$
chaotic signals. All of these are generated using  the standard tent
map, each of these begin  with a randomly chosen initial condition.
Each of these $M$ independent chaotic signals are to be of  length
$L$ each. She  then includes a $(M+1)^{st}$ member in this
collection which is also of a length $L$. This last signal is quite
arbitrary. It could be a message, some random noise, some periodic
signal or a chaotic signal generated by a source, either belonging
or not belonging to the family. The first $M$ signals naturally have
all their terms ranging from 0 to 1. For the last signal, an affine
transformation (which is obviously invertible) might be needed to
bring its terms within the same range.All the $M+1$ signals are now
treated as if they all belonged to the logistic (or the standard
tent map) family and each of them is transformed into a different
noise resistant map to once again form a separable cascade. The
trajectories from these are added and sent to the program simulating
the receiver. The receiver also receives the two parameters of the
affine transformation, representing scaling and shift. The receiver,
then  separates and reconstructs the signals into the standard tent
map setting. Based on the accuracy specified, the last $Q$ numbers
in the tail end of each signal are  abandoned. However, the $M+1$
sequences of the reduced length $L-Q$ each are highly accurately
reproduced by a single message of length $L$. Thus we have
$(M+1)\times(L-Q)$ numbers faithfully carried by L numbers. In our
example below $M$ is 20, $L$ is 350 and $Q$ is 50. So signal carried
information coding 18 times its length. Chaos is often described by
its random like character. If the signals were truly random, this
result would be impossible.

\section{Numerical verification}

\begin{figure}[!h]
\centering
\includegraphics[scale=0.4]{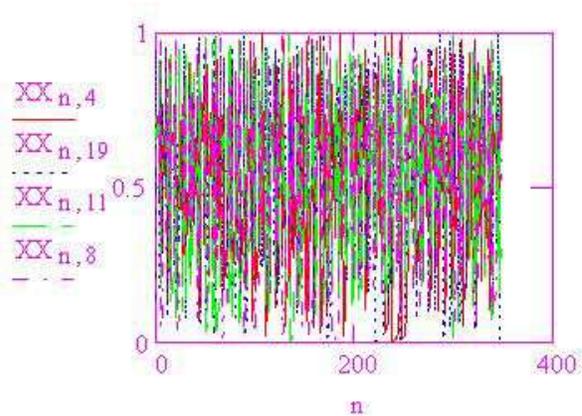}%
\caption[Figure 2]{The orbits of the 4 signals out of 20 chaotic
signals that will be mixed and separated. } \label{figure:fig2}
\end{figure}

\begin{figure}[!h]
\centering
\includegraphics[scale=0.4]{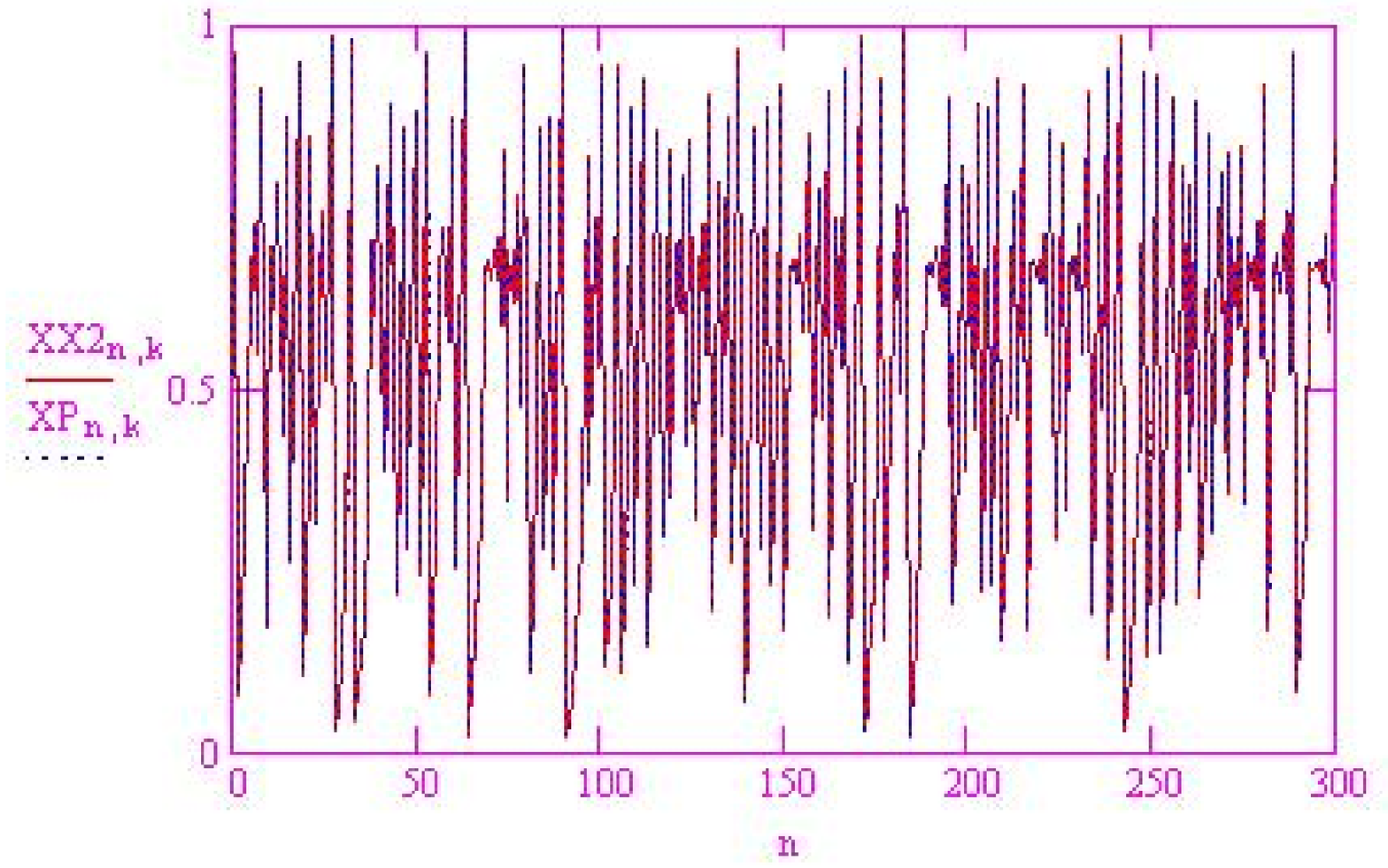}%
\includegraphics[scale=0.4]{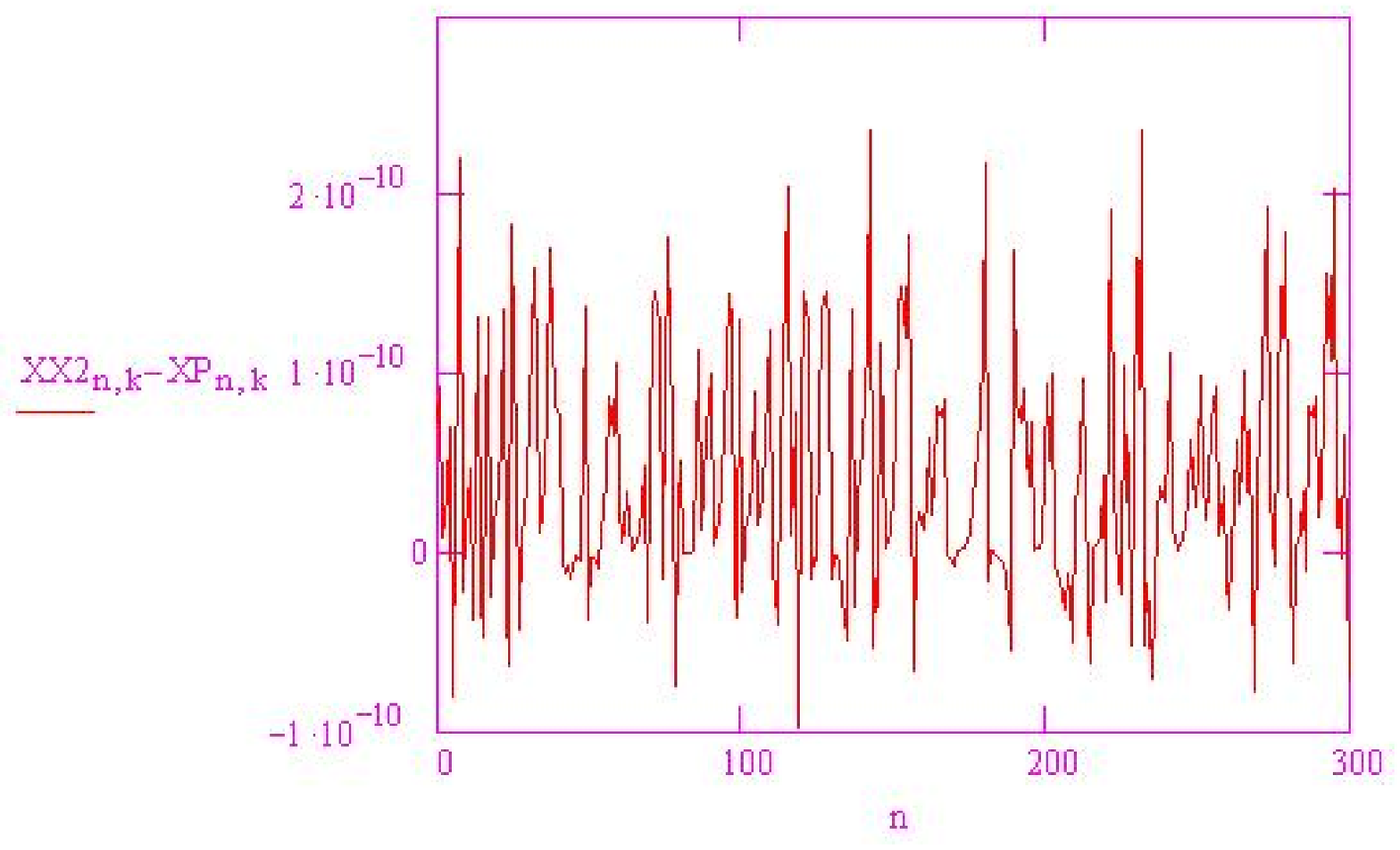}%
\caption[Figure 3]{(a) Left: The original (XX2) and the recovered
(XP) 20th chaotic signal. (b) Right: The difference of these two
signals.} \label{figure:fig3}
\end{figure}

\begin{figure}[!h]
\centering
\includegraphics[scale=0.4]{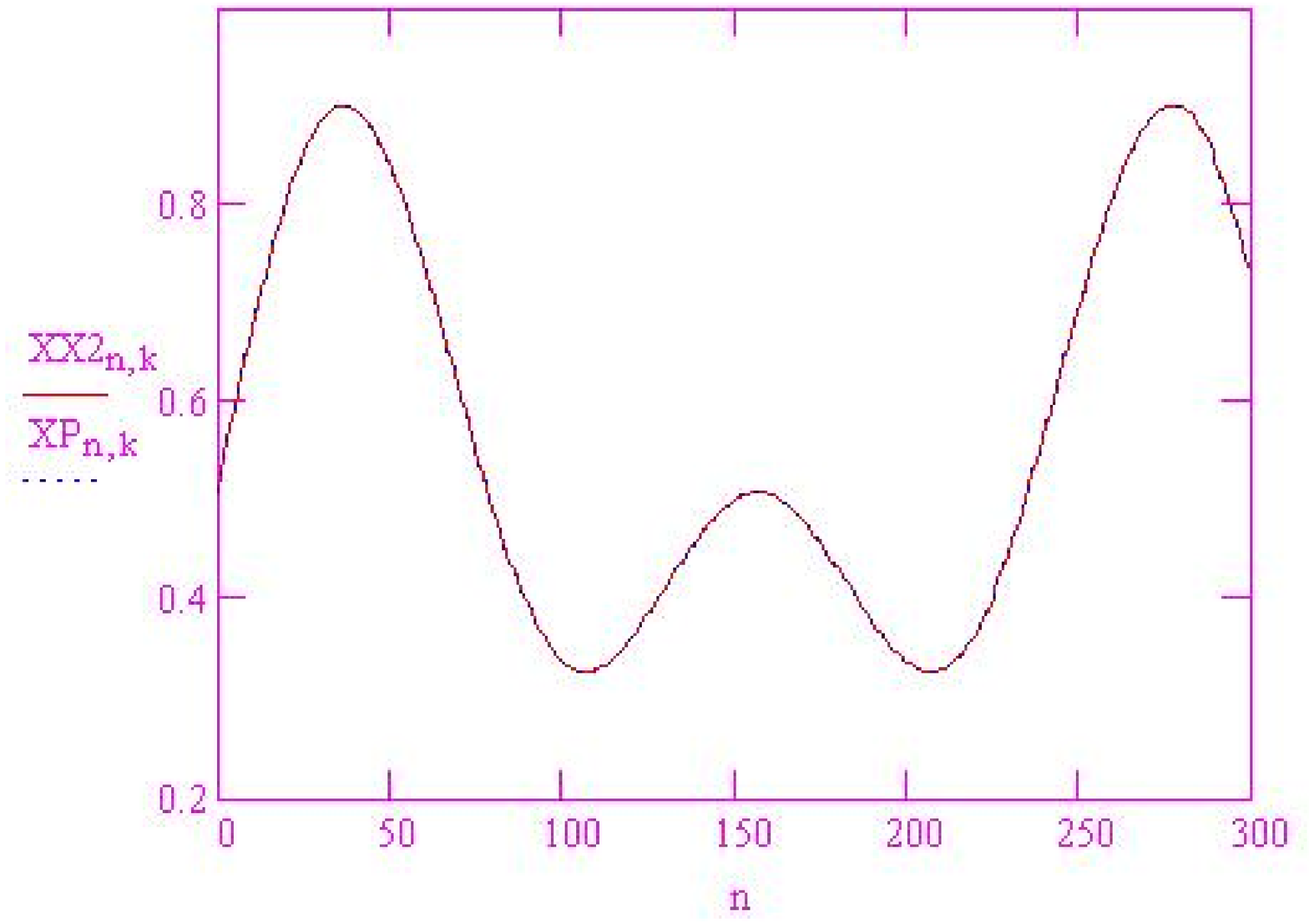}%
\includegraphics[scale=0.4]{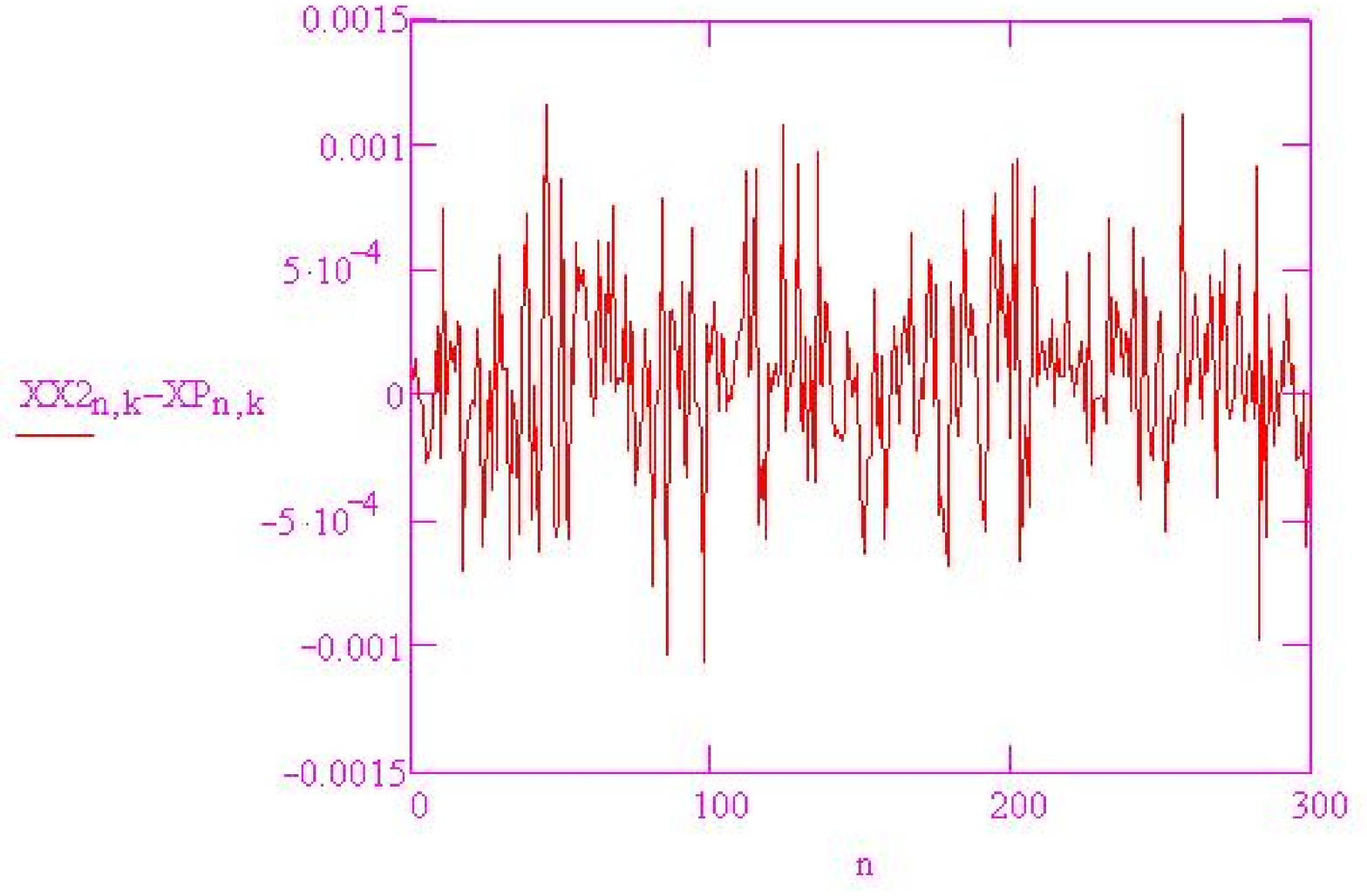}%
\caption[Figure 4]{(a) Left: The originally added periodic signal to
the mixture and its recovery. (b) Right: The difference between the
periodic signal and its recovery.} \label{figure:fig4}
\end{figure}

\begin{figure}[!h]
\centering
\includegraphics[scale=0.4]{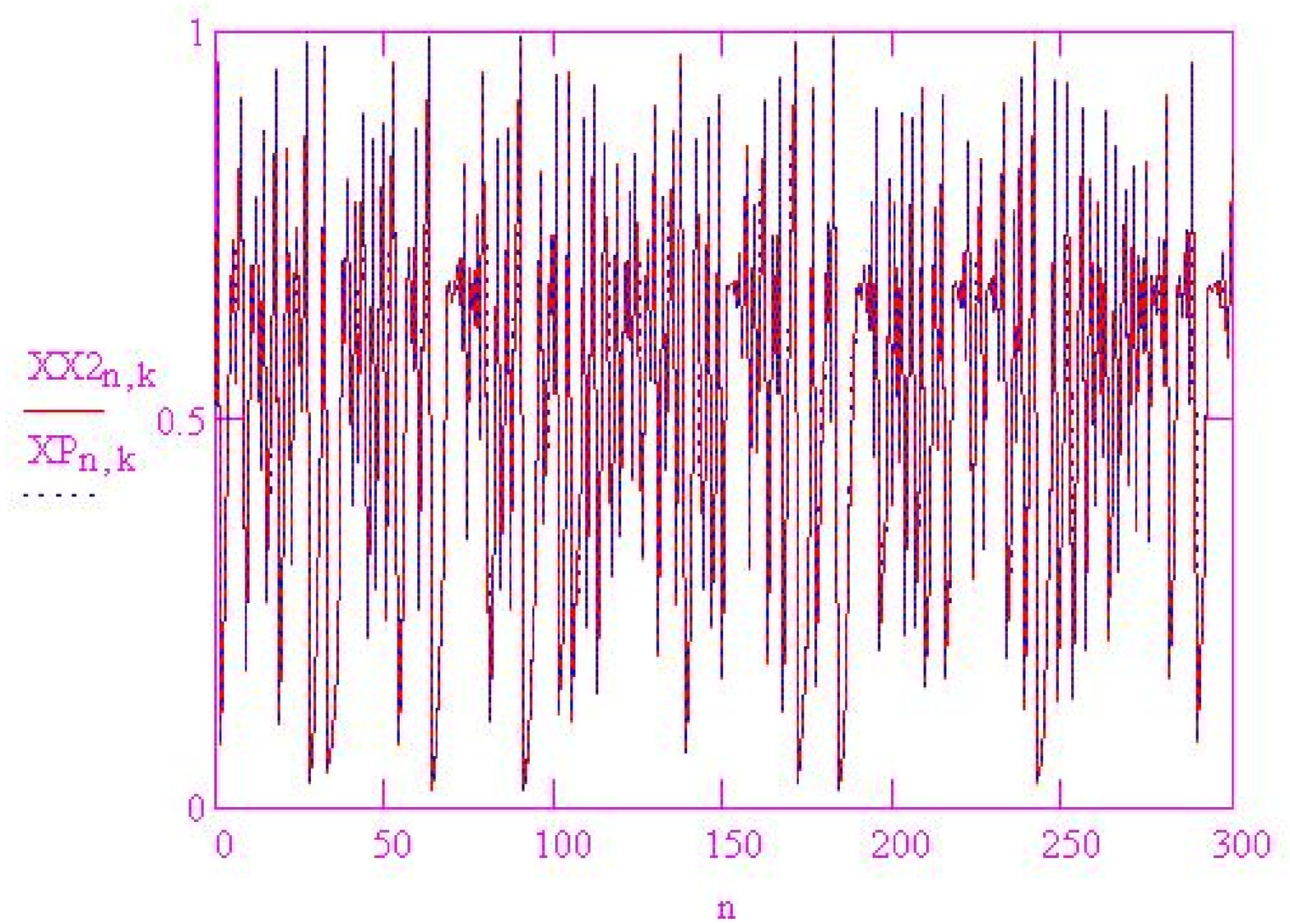}%
\includegraphics[scale=0.4]{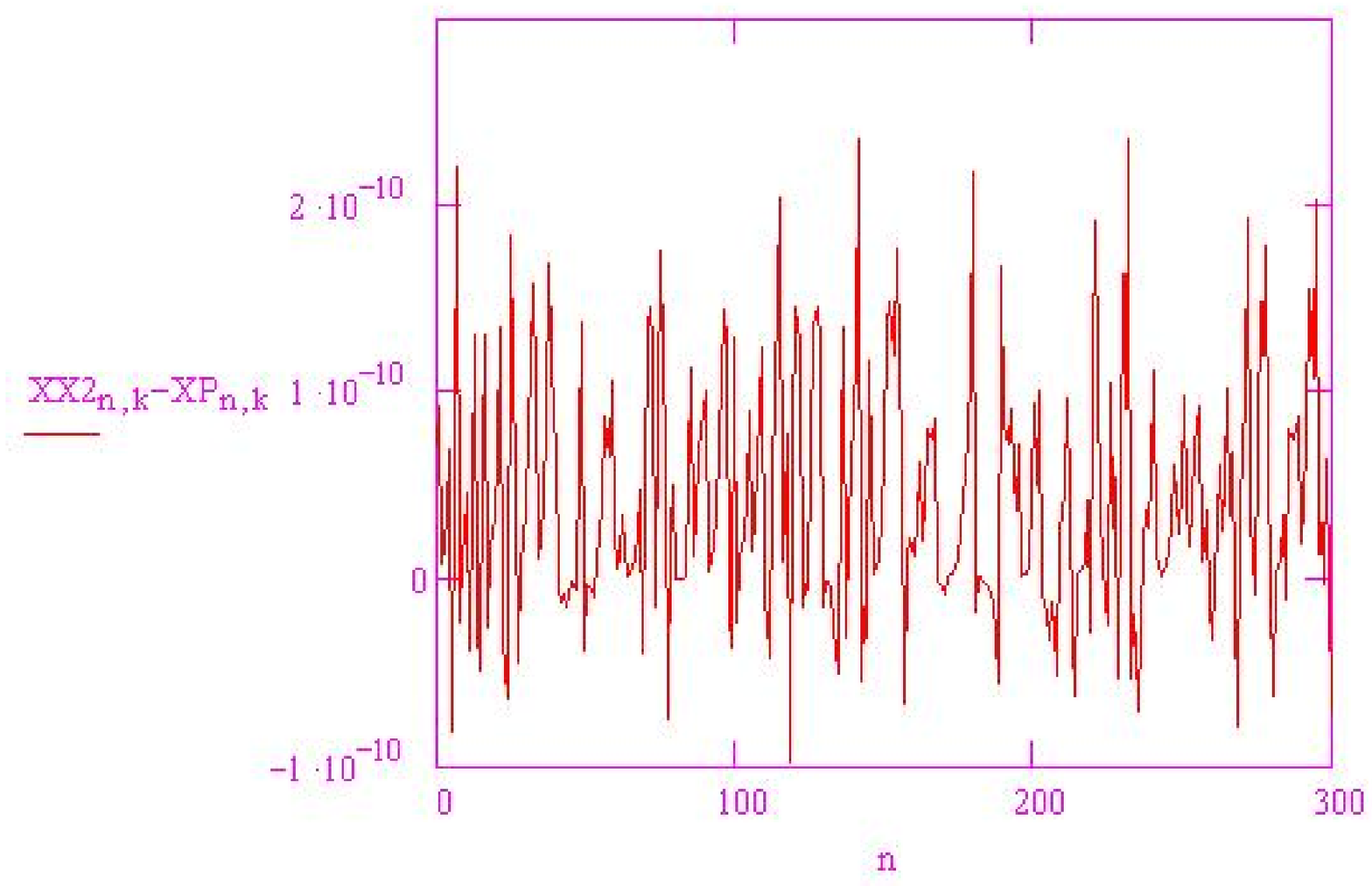}%
\caption[Figure 5]{(a) Left: The original (XX2) and the recovered
(XP) 20th chaotic signal for the case when added signal is random.
(b) Right: The difference between the 20th chaotic signal and its
recovery.} \label{figure:fig5}
\end{figure}

\begin{figure}[!h]
\centering
\includegraphics[scale=0.4]{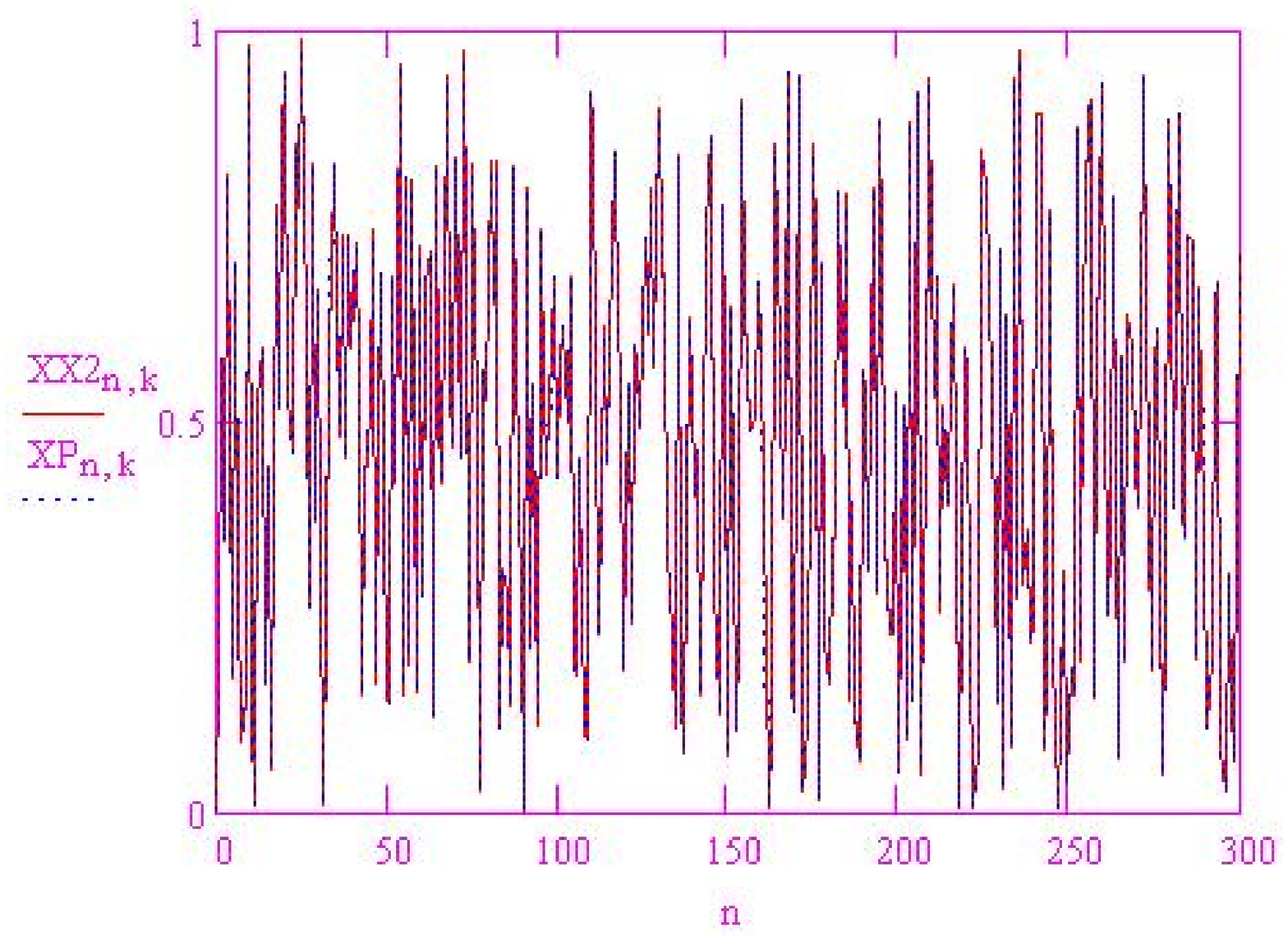}%
\includegraphics[scale=0.4]{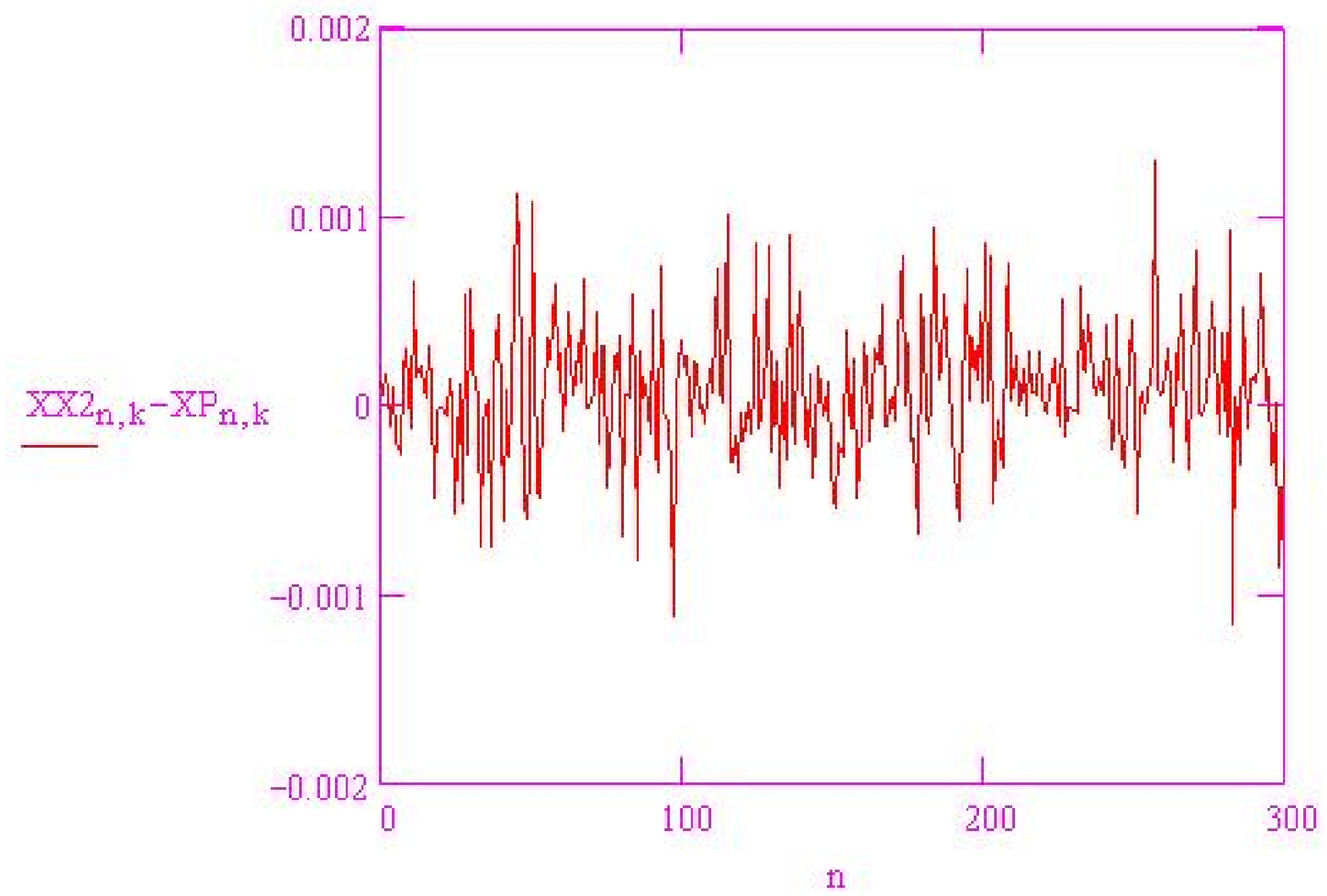}%
\caption[Figure 6]{(a) Left: The random signal which was added to
the mixture and its recovered signal. (b) Right: Difference between
the original and recovered random signal.} \label{figure:fig6}
\end{figure}

We chose $M$ =20 and $L$=350. In generating standard tent map
trajectory a modified algorithm was used (please see Appendix 1). We
show two extreme cases: 20 chaotic signals (with randomly chosen
initial conditions from the standard Tent map) were selected. Some
of these signals are seen in Figure~\ref{figure:fig2}. These were
added to a smooth periodic function which was scaled and a constant
was added so that all the terms were from 0 to 1. All these 21
signals were transformed and added. At the recovery stage, for the
last signal, the residue after predicting the first 20 signals was
chosen. Figures~\ref{figure:fig3}a and ~\ref{figure:fig3}b show the
results for the 20th chaotic signal which has the worst fit of all
the chaotic signals, yet the recovery is with an error of about
$10^{-10}$. Considering that the diameter of the data is close to 1,
this is very encouraging. Figures~\ref{figure:fig4}a and
~\ref{figure:fig4}b show the recovery of the periodic signal. The
error is of the order of $10^{-4}$. In the next case, another set of
20 chaotic signals are added to a signal which is fully random. In
this case, the 20th chaotic signal results
(Figures~\ref{figure:fig5}a, ~\ref{figure:fig5}b) are just as good
as the first case and the recovery of the random signal
(Figures~\ref{figure:fig6}a and ~\ref{figure:fig6}b) has a one order
more error than the periodic signal.

\section{Extending the possibilities using conjugacies}
 Now, let us consider a scenario which consists of a central mixing station which receives many
signals from possibly different maps, (although all belonging to the
same conjugate family, see Figure~\ref{figure:fig1}). At this
station, each of these signals are  transformed to the standard tent
map and mixed and then sent to a receiving station and the receiver
upon receiving follows a reversal procedure by which she recovers
all the signals with a remarkable accuracy. The transformations from
the standard tent map to one of its conjugates and their inverses
are quite straightforward and well established. However, once this
is established a fairly complex problem gets reduced to the one we
just simulated above. This opens up possibilities of cryptography,
coding and error correction and perhaps begins to explain how the
brain might use chaotic signals in its communication system.

\section{Conclusions}

It has been demonstrated that we can mix a large number of chaotic
signals and one completely arbitrary signal and  later a recipient
of this mixture   can separate each of these signals, one by one.
This has applications in cryptography and related areas. The paper
also helps further understand the nature of chaos.

\section*{Acknowledgments}

The author is very much indebted to Nithin Nagaraj for many fruitful
discussions and help in the preparation of the manuscript. The
origins of this paper can be traced to the inspiring lectures at
Indian Institute of Science by Professor V. Kannan of the Central
university of Hyderabad and subsequent discussions with him.

\section*{Appendix}

In generating standard tent map trajectory a modified algorithm was
used because the tent map has an attractor of 0 (when implemented on
a finite precision digital computer which stores and manipulates all
numbers in binary) for all initial conditions which can be expanded
in a finite number of terms in the binary representation.

The solution is quite simple: we first seek a topologically
conjugate map which just represents a scaling by a constant factor:
We begin with the standard tent map $T(x)$ which maps [0,1] to [0,1]
according to:

$$
= 2x,  ~~~  x \in [0,0.5R),
$$
$$
     = 2R-2x, ~~~ x \in [0.5R,R].
$$

To find a proper trajectory of the standard tent map,  the chosen
initial condition of the  map is multiplied by $R$. Then, one can
compute the orbit using the above equation. Once the orbit in the
scaled domain is computed, one just divides all the terms of the
orbit by $R$ again.

It is important that $R$ should not have a finite binary
representation also. When you compare the direct trajectory with the
modified trajectory, for the same initial condition, they both
remain close for a number of iterations depending on the machine
precision and then the conventional trajectory goes to zero. This is
because, a computer converts even a so called randomly chosen
initial condition into its approximation such that the approximated
initial condition has a finite binary representation.

\end{document}